\documentclass[a4paper]{jpconf}
\usepackage{graphicx}
\usepackage{amssymb}

\newcommand\pa[2]{\frac{\partial{#1}}{\partial{#2}}}
\def\Rs{R_{\odot}}
\def\Ru{R_{\textrm{\tiny top}}}
\begin{document}
\title{What do global p-modes tell us about banana cells?}

\author{Piyali Chatterjee}

\address{NORDITA, AlbaNova University Center, Roslagstullsbacken 23,
              SE 10691 Stockholm, Sweden}
              
\ead{piyalic@nordita.org}

\begin{abstract}
We have calculated the effects of giant convection cells also know as sectoral rolls or banana cells, on
p-mode splitting coefficients. We use the technique of quasi-degenerate perturbation theory formulated by Lavely \& Ritzwoller in order to estimate the frequency shifts. A possible way of detecting giant cells is to look for even splitting coefficients of  'nearly degenerate'  modes  
in the observational data since these modes have the largest shifts.
We find that banana cells having an azimuthal wave number of 16 and maximum vertical velocity 
of 180 m s$^{-1}$ cannot be ruled out from GONG data for even splitting coefficients.
\end{abstract}
\section{Introduction}
The power spectra of solar convective velocities show distinct peaks representing granules and 
supergranules but no distinct features at wavenumbers representative of mesogranules or 
giant cells (Wang 1989; 
Chou et al.~1991; Straus and Bonaccini 1997; Hathaway et al.~2000).
Numerical simulations of solar convection routinely show the existence of mesogranules and giant cells 
(Miesh et al.~2008, K\"apyl\"a et al.~2010). Of particular interest is the existence of sectoral rolls or
'banana cells' having maximum radial velocities of 200-300 ms$^{-1}$. The giant convective cells have always been elusive to observations at the
solar surface.
In the past there have been studies which have failed to detect giant cell motions 
(LaBonte, Howard and Gilman 1981; Snodgrass and Howard 1984; Chiang, Petro and Fonkal 1987
) as well as those hinting at their existence (Hathaway et al.~1996; Simon and Strous 1997).
With the availability of Dopplergrams from SOHO/MDI, it became possible to study such 
long-lived and large scale features more reliably.
Beck, Duvall and Scherrer (1998) were able to detect giant cells at the solar surface 
with large aspect ratio ($\sim 4$) and velocities $\sim 10$ m s$^{-1}$ using the MDI data. 
Hathaway et al.~(2000) used spherical harmonic spectra from full disk measurements to 
detect long-lived power at $l \le 64$.

Are banana cells observed regularly in direct numerical simulations (DNS), 
real or are they an artifact of insufficient resolution? 
Is it possible for techniques of global helioseismology to throw any light on the existence of giant cells?
 Traditionally, only degenerate perturbation theory (DPT) has been used to calculate the effect of 
rotation on the p-modes. But the first order contribution from poloidal flows 
(of which giant cells are a particular case) calculated
using the degenerate perturbation theory vanishes giving rise to the need to use the 
quasi degenerate perturbation theory (QDPT) which in contrast to DPT couples modes 
having slightly different unperturbed frequencies. 
Quasi degenerate perturbation theory  was applied to calculate shifts due to flows and asphericity in 
solar acoustic frequencies by Lavely and Ritzwoller (1992). 
Roth and Stix (1999, 2003) used QDPT to calculate the effect of giant cells on p-modes and claimed 
that giant cells could be found by modeling the asymmetries and line broadening in the 
solar power spectrum. They claimed that finite line width of the multiplets would limit the 
detection of the frequency splittings to vertical velocity amplitude of 100 m s$^{-1}$ or larger.
In Chatterjee and Antia (2009, CA09 now onwards), we used a different approach to examine if the use 
of quasi degenerate perturbation
theory introduces significant corrections in the frequency shifts over that obtained from degenerate
perturbation theory. We found the effect of rotation on the odd coefficients is negligible and 
hence using the degenerate theory is sufficient.
Additionally, CA09 also calculated the effect of N-S asymmetric component 
of rotation, single cell meridional circulation, giant cells and banana cells on p-modes. 
We also showed that for higher angular
degree of the poloidal flows, the 'nearly degenerate' modes would have very 
large frequency shifts (splitting coefficients). It is these splitting coefficients we can hope to 
detect in observational data.

In this paper we present results for banana cells but with an angular degree twice that used
in CA09. In Sect.~2, we present a brief description of the method 
for completeness. In Sect.~3 we present the results of our calculations and 
draw conclusions in Sect.~4.

\section{Perturbation of p- \& f-modes by giant cells}
Any perturbation calculation first requires definition of a base state with unperturbed eigenfrequencies 
as well as eigenvectors denoted $\omega_k$ and $\bf s_k$ respectively, 
which we shall call non-rotating spherically 
symmetric (NRSS) state. In this work 
we have used a standard solar model with the OPAL equation of
state (Rogers and Nayfonov 2002) and OPAL opacities 
(Iglesias and Rogers 1996) and use the formulation due
to Canuto and Mazitelli (1991) to calculate the convective flux.  The equations of motion for a mode $k$ with eigenfrequency $\omega_k$ for a NRSS model and a model perturbed by addition of differential rotation and/or large scale flow can be respectively represented by,
\begin{eqnarray}
\label{eq:qdpt1}
{\mathcal{L}_0}{\bf s_k}&=&-\rho_0\omega_k^2 {\bf s_k},\\
\label{eq:qdpt2}
{\mathcal{L}_0}{\bf s'_k}+{\mathcal{L}_1}{\bf s'_k}&=&-\rho_0{\omega'_k}^2 {\bf s'_k},
\end{eqnarray}
where $\omega_k'$ is the perturbed frequency and 
${\bf s'_k}= \sum_{k'\in K} a_{k'} {\bf s_{k'}}$  is the perturbed eigenvector.
Taking scalar product with ${\bf s_j}$ in equation~(\ref{eq:qdpt2}) and using the 
notation ${\mathcal H}_{jk'} = - \int {\bf s_{j}}^{\dagger} {\mathcal L}_1 {\bf s_{k'}} dV$ 
and the definition  $\mathcal{L}_1 {\bf s_k} = -2i\omega_{\textrm{\scriptsize ref}}\rho_0({\bf v.\nabla)s_k}$, we obtain the matrix eigenvalue equation,
\begin{equation}
\label{eq:qdpt3}
 \sum_{k' \in K} \left\lbrace {\mathcal H}_{jk'} + \delta_{k'j} (\omega_{k'}^2 - \omega_{\textrm{ref}}^2)\right\rbrace a_{k'} =  ({\omega'_k}^2-\omega_{\textrm{ref}}^2)a_{j},
\end{equation}
with eigenvalue $\lambda = ({\omega'_k}^2- \omega_{\textrm{ref}}^2)$ and eigenvector $X_{j} = \left\lbrace a_{j} \right\rbrace$.
Here $\omega_\textrm{ref}$ is a reference frequency which approximates
$\omega_k'$. In this work we use $\omega_\textrm{ref}=\omega_k$, the frequency
of the mode being perturbed.
For details on calculation of the matrix elements, the reader is referred to 
Sect. 3.2 of CA 09. Essentially we use the Wigner-Eckart theorem 
(equation~5.4.1 of Edmonds 1960) which states that the general matrix element 
of any tensor perturbation operator can be expanded in terms of Wigner 3$j$ symbols 
whose coefficients of expansion are independent of azimuthal order $m$ and $m'$. 

An important thing to remember is that for two modes with 
frequencies $\omega_1$ and $\omega_2$, the frequency shift is given by,
\begin{equation}
\label{eq:del}
\delta\nu = \frac{\omega'_2 -\omega_2}{2\pi} \sim \frac{{\mathcal{H}}_{12}^2}{4\pi\omega_2\Delta}\;
\end{equation}
where ${\mathcal{H}}_{12}$ is the coupling matrix between the two modes 
and $\Delta = ({\omega_{2}}^2- \omega_{1}^2)$. Traditionally in 
helioseismology, the frequency shift as a function of $m$ is described in terms of 
splitting
coefficients for all modes that are detected. 
These coefficients are defined
by (e.g., Ritzwoller and Lavely 1991)
\begin{equation}
\label{eq:m}
\omega_{nlm} = \omega_{nl} + \sum_{q=0}a_q^{(nl)}\mathcal{P}^{l}_q(m),
\end{equation}
where $\omega_{nl}$ is the mean frequency of the multiplet,
$\mathcal{P}_q^{l}(m)$ are the orthogonal polynomials of degree $q$ and $a_q^{(nl)}$'s are the so called splitting coefficients. A multiplet with frequency $\omega_{nl}$ is $2l+1$-fold degenerate 
in absence of rotation, magnetic field, poloidal flows and asphericity. 
Following earlier works (Lavely and Ritzwoller 1992; Roth, Howe and Komm 2002) we express the velocity field in terms of spherical harmonics. For completeness, we give the expression here again.
\begin{eqnarray}
\label{eq:flow}
\nonumber
{\bf v}(r,\theta,\phi) &=& \mathrm{Re}[u_s^t(r) Y_s^t(\theta, \phi) {\bf{\hat{r}}} + v_s^t(r) {\bf \nabla}_{h}Y_s^t(\theta, \phi) - \\
&&\qquad w_s^t(r) {\bf{\hat{r}}}\times {\bf\nabla}_{h}Y_s^t(\theta, \phi)] .
\end{eqnarray}
The quantities $u_s^t, v_s^t$ and $w_s^t$ determine the radial profiles of the flows and ${\bf\nabla}_{h}$ is the horizontal gradient operator. The $\mathrm{Re}$ refers to using only the real part of the spherical harmonics as in,
\begin{equation}
\mathrm{Re}[Y_s^t(\theta,\phi)]=\begin{cases}{[Y_s^{-t}(\theta,\phi)+Y_s^{t}(\theta,\phi)]/2 & if $t$ is even, \cr 
\noalign{\smallskip}
[Y_s^{-t}(\theta,\phi)-Y_s^{t}(\theta,\phi)]/2 & if $t$ is odd.\cr}\end{cases}
\end{equation}
The first two terms in equation~(\ref{eq:flow}) define the poloidal component of the flow whereas the last term is the toroidal component.
By the poloidal component, we imply the meridional and non-zonal toroidal flows (average over $\phi$ direction is zero) e.g., (i) the meridional circulation which carries mass poleward near the surface and sinks near the poles and (ii) the giant convection cells, respectively.
These flows are also called large scale flows to distinguish them from other small scale flows like the turbulent eddies which are of the size smaller than
the typical scale of global modes used in helioseismology.

The sectoral rolls or banana cells are characterised by $s=t$. 
CA09 used $s =t=8$ for banana cells. Visual inspection
of the snapshots of vertical velocity in Fig.~1 of Miesch et al. (2006) as well as 
Fig.~2c of K\"apyl\"a et al. (2010) reveals that banana cells observed in DNS of
stellar convection has $s\sim 16, t\sim 16$. In addition Miesch et al. (2008) also put 
the value of the maximum radial velocity at $200$ m s$^{-1}$. So it makes sense to repeat our calculations for a flow with an angular dependence $Y_{16}^{16}(\theta, \phi)$. In this calculation we have 
used only the estimate of the magnitude of maximum radial velocities and their angular degree from the DNS. 
The solar structure we use for frequency shift calculations come from a spherically 
symmetric non-rotating  standard solar model (NRSS model) and the rotation splittings needed are obtained 
from temporally averaged GONG data. These two quantities are
very different for DNS since they have much shallower density stratification and sometimes give a rotation
profile which is constant on cylinders rather than cones as for the Sun. 
A more realistic calculation would be to use the stratification and the rotation rate from the DNS to
define the unperturbed state. Nevertheless, a first calculation using a NRSS model and GONG rotation profile
can provide valuable insights into the possibility of such poloidal flows.

In presence of only the poloidal flow ($w_s^t = 0$) we can apply the equation of mass conservation
$\nabla.(\rho_{0}{\bf v}) = 0$ to get a relation between $u_s^t(r)$ and $v_s^t(r)$ e.g.,
\begin{equation}
\label{eq:vst}
v_s^t(r) = \frac{1}{r}\pa{}{r} \left[\frac{\rho_{0} r^2 u_s^t(r)}{s(s+1)}\right]\;.
\end{equation}
Here $\rho_0(r)$ is the density in a spherically symmetric solar model. So now it only remains to choose $u_s^t(r)$ appropriately and $v_s^t(r)$ will be determined by equation~(\ref{eq:vst}). We choose the radial profile of $u_s^t(r)$ as given by equation~(19) of Roth, Howe \& Komm (2002): 

\begin{equation}
\label{eq:profile}
u_s^t(r)=\cases{u_0\frac{4(\Ru-r)(r-r_b)}{(\Ru-r_b)^2} & if $r_b\le r\le \Ru$, \cr 0 & otherwise.\cr}
\end{equation}
Here $r_b = 0.7\Rs$ and $\Ru = \Rs$ define the boundaries of region where the flow is
confined. The coupling matrix  in Eq.~(\ref{eq:qdpt3}) 
calculated for the flow defined above is proportional to $u_0$. Hence using Eq.~(\ref{eq:del}), we have
the frequency shift, $\delta\nu \propto u_0^2$.

\section{Results}

As explained in CA09, this calculation is little more involved 
as a non-zero $t$ allows coupling between different $m$ and $m'$ of the p-modes.
According to the selection rules imposed by the Wigner 3$j$ symbols, the
$s=16,t=16$ flow couples the mode $(n,l,m)$ with $(n',l',m\pm 16)$. Hence it becomes important to 
take into account the 
effect of rotational splitting on $\Delta$ before calculating the effect of these kind of poloidal 
flows. Thus, the difference 
of the square of frequencies, $\Delta$ is no longer independent of $m$. 
The couplings within a multiplet i.e.,  $(n,l,m) \rightleftarrows (n,l,m \pm 16)$ 
happen to be zero because of the anti-symmetry of the matrix elements (see Eqs.~(16)-(18) in CA09). 

In the upper 
panel of Fig.~\ref{fig:n10l46}, we show 
the frequency shift of the mode with $(n, l) = (11, 38)$
due to coupling with $(10, 46)$. 
From Eq.~(\ref{eq:del}), we have the frequency shift, $\delta \nu \propto 1 / \Delta$. 
The $\delta \nu (m)$ shows a discontinuity at $m=0$ since 
$\Delta$ for the $(11, 38, m) \rightleftarrows (10, 46, m-16)$ becomes zero at $m=0$ as 
shown in the lower panel. 
The reader is encouraged to compare this figure with Fig.~7 of 
CA09 where they plot similar curves for coupling of the 
modes $(18, 61, m) \rightleftarrows (17, 69, m\pm 8)$.
It is these discontinuities which give rise to a large value of splitting coefficients, $a_q$.
Also notice the asymmetry of $\delta \nu (m)$ about $m=0$ in Fig.~\ref{fig:n10l46}. 
This means that not only the even coefficients, $a_{2q}$ 
but also the odd coefficients, $a_{2q+1}$ are non-zero. 
Usual inversion procedure for rotation neglects giant cells and assumes
that the odd splitting coefficients arise only from rotation. If there
are additional contributions to these coefficients, rotation inversions
may not give correct results.  In CA09, we performed a 1.5d rotation inversion on odd coefficients 
$a_1$ and $a_3$ to detect any discernible feature in the inverted profile. 
However, since the magnitude of the features in the inverted profile are smaller than the inversion 
errors we concluded that giant cells with $u_0 \le 100$ m s$^{-1}$ do not have 
much effect on the rotation inversion.
The splitting coefficients $a_1, a_2$ and $a_3$ as a function of lower turning point radius 
and normalised by corresponding errors in the observational data from GONG are 
shown in Fig.~\ref{fig:a1a2a3}. One can easily compare these coefficients for $s=8, t=8$ 
and $s=16, t=16$ flows. While the $Y_8^8(\theta, \phi)$ flow with an amplitude of 
$u_0=100$ m s$^{-1}$  already produces a maximum $a_2$ which is twice the observational error, a
$Y_{16}^{16}$ flow with an amplitude  $u_0=300$ m s$^{-1}$ produces a maximum $a_2$ which is half
the observational error. A similar thing may be said about the coefficients $a_1$ and $a_3$ as well. 
However, we have not performed a 1.5 D inversion for the $s=16, t=16$ case. We do not see a clear signal 
of 'nearly degenerate' modes in the observational data. Hence, like in CA09, 
in order put an upper limit
on the amplitude of the flows we compare the $a_2$ values from 
theory with observational errors to calculate the confidence level of the upper limit. 
For the $Y_{16}^{16}(\theta, \phi)$, the confidence limit estimated by 
maximum value of $a_2 /\sigma_2$ is less than 0.6 for a flow with an 
amplitude $u_0=300$m s$^{-1}$ for several GONG data sets. In terms of the maximum radial velocity, 
$u_r^{\rm max}\sim 180$m s$^{-1}$. Thus banana cells with such amplitudes cannot be ruled 
out using the GONG data sets used. Another interesting observation from 
Fig.~\ref{fig:a1a2a3} is the large values of $a_{q}$ at $r_t \sim 0.7\Rs$ for both $t=8$ and $16$. 
We have checked that this is not due to $r_b = 0.7\Rs$, but is an intrinsic property of the solar model 
used. The rotation profile used to calculate the rotational splittings for $t\neq 0$ giant cell flows
may also be responsible. Nevertheless, since we also see this behaviour for $s=8, t=0$
flow (see Fig.~5 of CA09), we may conclude that the modes which have their turning points 
near the base of the convection zone are coupled strongly due to such poloidal flows.
\begin{figure} 
\label{fig:n10l46}
\centering{\includegraphics[width=.7\textwidth]{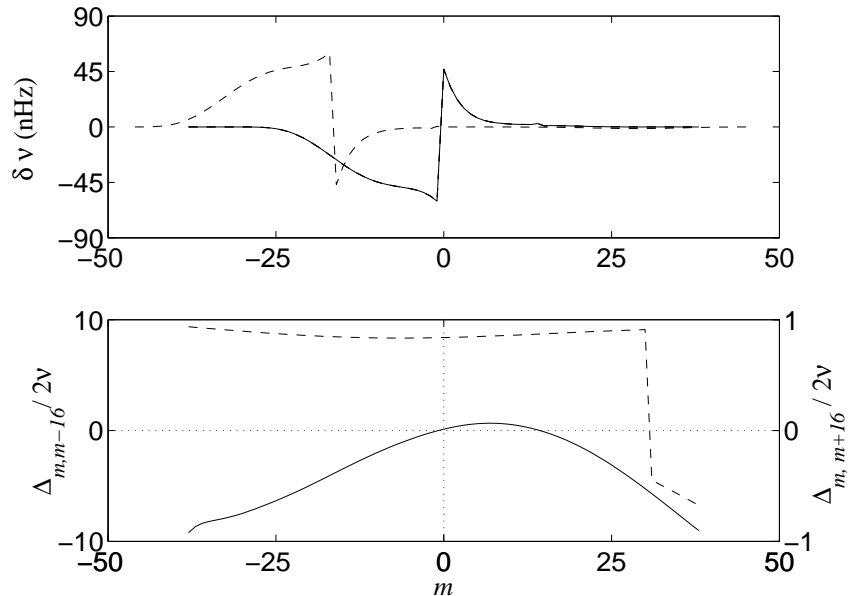}}
\caption{\label{fig:n10l46} Upper panel: The {\em solid line} gives the frequency shift 
because of interaction between modes $(11, 38, m)$ having lower turning radius $r_t = 0.69\Rs$ 
and $(10, 46, m \pm 16)$ having $r_t = 0.74\Rs$,
due to the flow with an angular dependence $Y_{16}^{16}(\theta,\phi)$. 
The frequency shift for the mode $(10, 46)$ is given by the {\em dashed line}.
Lower panel: $\Delta_{m,m-16}/2\nu$ ({\em solid line}) for the coupling 
$(11,38,m)\rightleftarrows(10,46,m-16)$; and $\Delta_{m,m+16}/2\nu$ ({\em dashed line}) 
for the coupling $(11,38,m)\rightleftarrows(10,46,m+16)$. These values are in $\mu$Hz.}
\end{figure}

\begin{figure} 
\label{fig:a1a2a3}
\includegraphics[width=.7\textwidth]{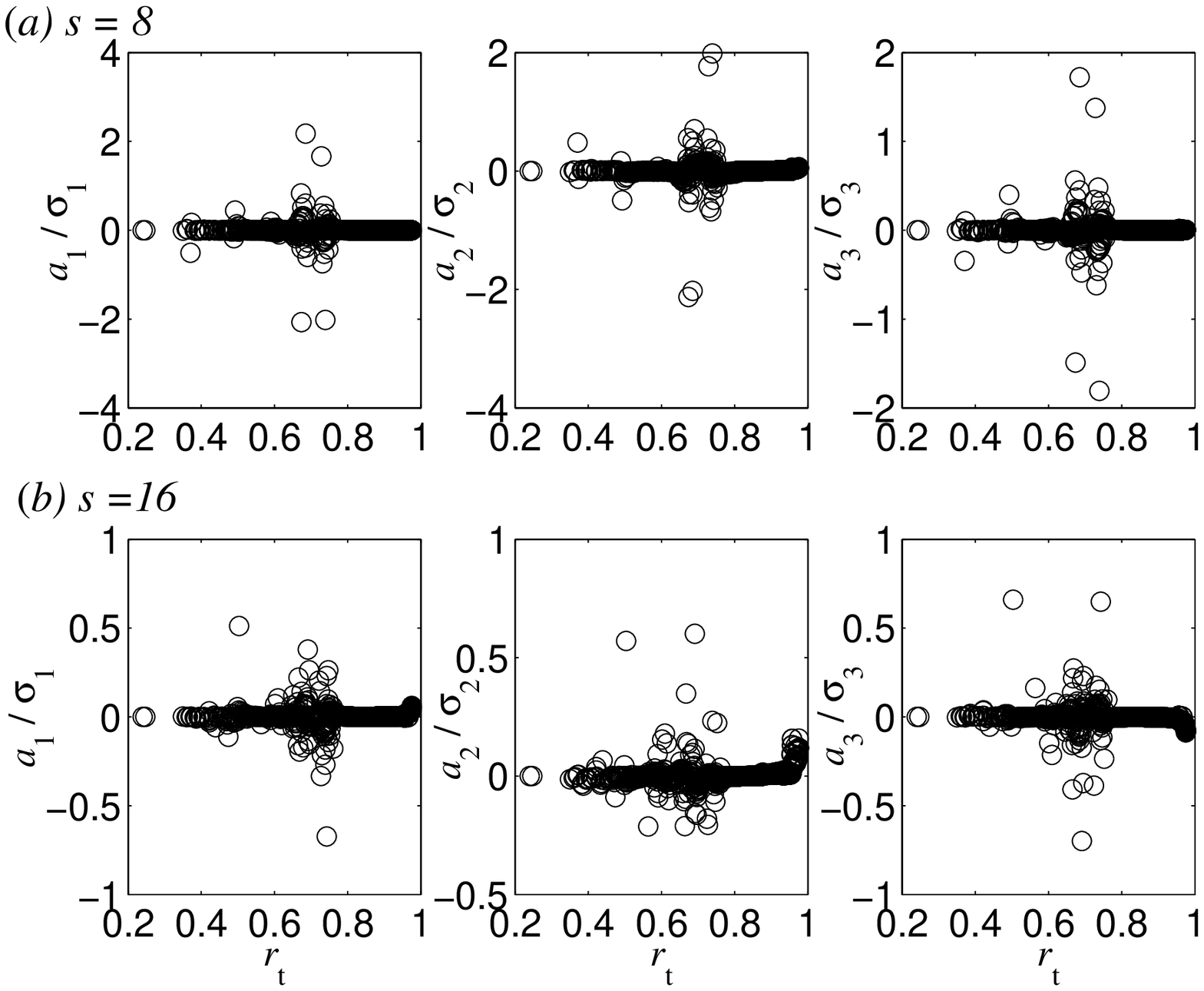}
\caption{\label{fig:a1a2a3} (a) $a_1/\sigma_1$, $a_2/\sigma_2$ and $a_3/\sigma_2$ as a function of lower turning point radius $r_t$ for the $Y_{8}^{8}(\theta, \phi)$ kind of flow with $u_0 =100$ m s$^{-1}$. 
(b) Same as above but for a flow varying as $Y_{16}^{16}(\theta, \phi)$ with 
$u_0=300$ m s$^{-1}$. The $\sigma_1$, $\sigma_2$ and $\sigma_3$ are the errors 
in the corresponding observational splitting coefficients $a_1$, $a_2$
and $a_3$ from the GONG data set centered about November 2002.}
\end{figure}
\section{Conclusions} 
We have calculated the effects of giant cells with an azimuthal degree $16$ on p-mode 
splitting coefficients. We use the technique outlined in Chatterjee \& Antia (2009) to 
perform this calculation. We find that from observations we cannot rule out the existence 
of giant cells with an angular
variation $Y_{16}^{16}(\theta, \phi)$ and an amplitude of $u_0=300$ m s$^{-1}$ 
since the splitting
coefficients are less than the observational errors. It is important to remember that $a_q \propto u_0^2$.
In other words we can say that a flow varying as  $Y_{8}^{8}(\theta, \phi)$ 
in the Sun can be ruled out with a confidence level estimated by the maximum value of the ratio 
$a_2 / \sigma_2\sim 2$ which is 30 times more
than the confidence level for a flow varying as $Y_{16}^{16}(\theta, \phi)$ 
with the same amplitude $u_0=100$ m s$^{-1}$. 
Speaking in terms of maximum vertical velocity, $u_r$, according to 
Eq.~(\ref{eq:flow}), a 
$u_0=100$ m s$^{-1}$ for a $s=8, t=8$ flow 
corresponds to $u_r^{\rm max} = 50$ m s$^{-1}$  whereas for a $s=16, t=16$ flow implies
a  $u_r^{\rm max} = 60$ m s$^{-1}$. 

We do not find any evidence from global helioseismology 
which may point at the appearance of banana cells with an azimuthal degree $t=16$ 
in the DNS to being just an artifact of insufficient grid resolution. But we believe that in 
the DNS the velocity spectra would  peak at the wavenumbers of the banana cells since 
they are so visually conspicuous in contrast to observations where there is no clear peak 
at giant cell scales. This may be because we do not resolve very many scales smaller than the
banana cells in the DNS. But it is important to remember that we have used very 
regular analytical expressions for the banana cells. In the Sun, the cells will not only be irregular 
but also have finite lifetimes. The GONG data sets are averaged over 108 days and this 
may be somewhat longer than the lifetimes of giant cells and hence
the signal may be averaged out. We can look at shorter data sets, but in
that case the errors would be larger.

\section*{Acknowledgements}
We thank the organisers of the GONG 2010 SOHO-24 meeting in Aix-en-Provence 
for an excellent meeting.
This work  utilizes data obtained by the Global Oscillation
Network Group (GONG) project, managed by the National Solar Observatory,
which is
operated by AURA, Inc. under a cooperative agreement with the
National Science Foundation. The data were acquired by instruments
operated by the Big Bear Solar Observatory, High Altitude Observatory,
Learmonth Solar Observatory, Udaipur Solar Observatory, Instituto de
Astrofisico de Canarias, and Cerro Tololo Inter-American Observatory.

\end{document}